\documentclass[useAMS,usenatbib]{mn2e}%
\usepackage{graphicx}
\usepackage{amsmath}
\usepackage{amsfonts}
\usepackage{amssymb}%
\newcommand\apj{{ApJ}}%
\newcommand\apjl{{ApJ}}%
\newcommand\aap{{A\&A}}%
\newcommand\mnras{{MNRAS}}%
\newcommand\prd{{Phys.~Rev.~D}}%
\begin{document}

\title[Transparency of electron-positron-photon plasma]
  {Transparency of an instantaneously created electron-positron-photon plasma}
\author[B\'egu\'e and Vereshchagin]
  {D.~ B\'egu\'e,${^1}{^2}{^3}$
  G. V. Vereshchagin${^1}{^2}$
 \\
  $^1$University of Roma ``Sapienza'', 00185, p.le A. Moro 5, Rome, Italy\\
  $^2$ICRANet, 65122, p.le della Repubblica, 10, Pescara Italy\\
  $^3$Erasmus Mundus Joint Doctorate IRAP PhD student}
\date{}

\maketitle

\begin{abstract}

The problem of the expansion of a relativistic plasma generated when a
large amount of energy is released in a small volume has been considered by many authors. We use the analytical solution of \cite{bisno} for the spherically symmetric relativistic expansion.

The light-curves and the spectra from transparency of an electron-positron-photon plasma are obtained. We compare our results with the work of \cite{goodman}.

\end{abstract}


\section{Introduction}

The problem of the release of a large amount of energy in a small volume has been considered for the first time by \cite{fermi}, who proposed a statistical theory for computing high energy collisions of protons with multiple production of particles. \cite{landau} noticed that the initial expansion of the system can be treated within relativistic hydrodynamics. Due to highly relativistic velocities of the colliding particles, the region of collision appears to be highly contracted in one direction. Consequently the problem can be reduced to one dimensional relativistic hydrodynamics in plane geometry. Landau found an approximate solution of this problem. Later on, an exact solution has been given by \cite{khalatnikov}.

A similar problem has been considered in application to Gamma Ray Bursts (GRBs) within spherical geometry. \cite{goodman} considered the
fate of a large quantity of energy in photons and electron-positron pairs, initially confined to a sphere in equilibrium at temperature above MeV, and then allowed to expand freely. He solved numerically the relativistic hydrodynamics equations.
He found that the plasma expands and cools down to non relativistic temperatures. Then due to the exponential dependence of pairs density on temperature and consequently on radius, the system becomes transparent suddenly.
He computed the energy distribution of the photon flux received by a distant observer by integrating over the volume of the system at the moment of transparency. The spectrum was found to be nearly thermal.

An approximate analytic solution for the problem of relativistic spherical expansion into vacuum of an instantly created ultra-relativistic
plasma has been given by \cite{bisno}. In this paper we used this solution in order to find the observed spectra from transparency of
electron-proton-photon plasma. The problem is intrinsically dynamic with the photosphere evolving rapidly with time. The only method available to compute the photospheric emission in such dynamical case is the one by \cite{RSV}. This method solves the radiative transfer equation assuming the source function to be isotropic and  thermal.

The applications of the results presented in our paper are twofold. Firstly in the case of very low baryon contamination, it is a
natural extension of \cite{RSV} who considered finite wind profiles. It also finds direct application in the interpretation of
the spectra of some GRBs in the context of the Fireshell model, see e.g. \cite{ruffini07} and references therein. Indeed, \cite{muccino} analysed the short GRB090227B and interpreted the thermal first episode as an almost pure electron-positron-photon plasma reaching transparency. One has to keep in mind that the photospheric component of long duration (on the order of seconds) seen in some bursts cannot be explained within this model. It is usually interpreted within the relativistic wind model of \cite{Paczynski1990}.

In section 2 we introduce the solution from \cite{bisno}, section 3 gives the method to compute the light-curves and the spectra. In section 4, the numerical results are presented. Discussion and conclusion follow.

\section{Approximate solution of the relativistic hydrodynamic equations}

\begin{figure}
\centering
\includegraphics[scale=0.70]{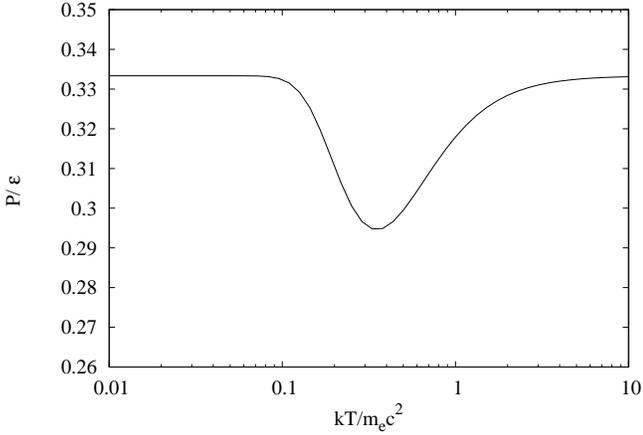}
\caption{Equation of state of the optically thick electron-positron-photon plasma as a function of the co-moving temperature $T$.}
\label{Fig:eqs}
\end{figure}

To obtain a realistic profile of the shell, we used the solution of \cite{bisno}, for details see Appendix, that is valid for the ultra-relativistic equation of
state $\epsilon=3P$, where $\epsilon$ is the co-moving energy density and $P$ is the pressure, and ultra-relativistic expansion velocity
$v \approx c$. For an optically thick system of electron-positron pairs and photons, all particles give a contribution to the equation of state.
Assuming thermal equilibrium, one can find the equation of state in this system. It is presented in Fig.\ref{Fig:eqs} for different values of
the dimensional co-moving temperature $T$.

At high temperatures $k_BT \gg m_ec^2$, where $k_B$ is the Boltzmann constant, $m_e$ the electron mass and $c$
the speed of light, the equation of state is ultra-relativistic.
At non-relativistic temperatures ($k_BT \ll m_ec^2$), the equation of state is also ultra-relativistic since the contribution of
non-relativistic electron-positron pairs is small (see also \cite{goodman}). As can be seen from Fig.\ref{Fig:eqs} the maximum deviation
from the value $1/3$ is achieved at the temperature $k_BT=0.33m_ec^2$ and it amounts for 12\%. This fact justifies the ultra-relativistic
equation of state in the optically thick electron-positron-photon plasma. It is also known that such plasma, being optically thick,
expands with acceleration and reaches ultra-relativistic velocity of expansion before becoming transparent (c.f. \cite{goodman}). Consequently the
solution of \cite{bisno} can be applied to this system.

\begin{figure}
\centering
\includegraphics[scale=0.45]{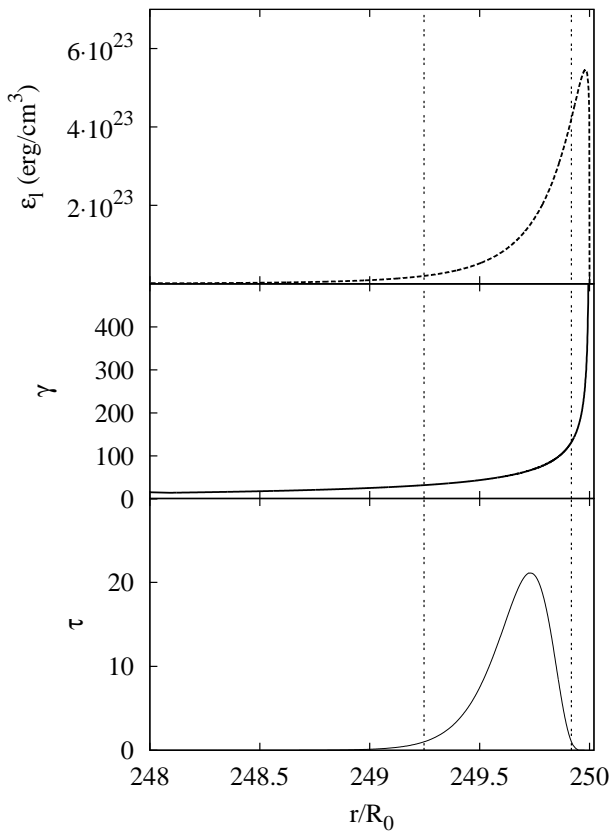}
\caption{Laboratory energy density (top), Lorentz factor (middle) and optical depth (bottom) radial profiles at $t=250 R_0/c$. The two vertical lines represent the radii at which $\tau(t,r,\phi=0)=1$. The main part of the energy has already been emitted, as can be seen from the top panel.}
\label{Fig:profil}
\end{figure}

The analytical solution of the relativistic hydrodynamic equations given by \cite{bisno} allows to compute the Lorentz factor $\Gamma$ and co-moving
energy density $\epsilon$ at arbitrary laboratory time $t$ and laboratory radius $r$:
\begin{align}
\epsilon (t,r)& =\frac{2\epsilon_{10}g_0 f(t,r) \tilde{f}(t,r)}{\xi r^3} \label{Eq:epsilon}\text{,}\\
\Gamma (t,r)& = \sqrt{\frac{rf(t,r)}{2 \xi \tilde{f}(t,r)}} \label{Eq:gammaeq} \text{,}
\end{align}
where $f$ and $\tilde{f}$ are given in Appendix \ref{appendixA}, $\xi=c t-r$ measures the depth within the shell with $g_0$ and $\epsilon_{10}$ being parameters
of the solution\footnote{The formula (3.21) for the Lorentz factor in the \cite{bisno} paper contains a misprint.}. For large $t$, the solution
describes a thin shell with constant laboratory width (see \cite{piran93,ruffini99,ruffini00}) propagating radially with $\Gamma \gg1$, see Fig.\ref{Fig:profil}.

From the co-moving energy density, the co-moving temperature can be found as $T(t,r)=(c\epsilon(t,r) / (4\sigma_{SB}))^{1/4}$, where
$\sigma_{SB}$ is the Stefan-Boltzmann constant. Electron-positron-photon plasma with macroscopic size, gets transparent when the
temperature decreases to the value $k_BT\sim0.04 m_e c^2$ \citep{RSV}. At such non-relativistic temperature, the co-moving
number density of pairs is given by  (e.g. \cite{douglas}):
\begin{align}
n_c^{\pm}(t,r)=4\left ( \frac{m_e}{h}\right )^3 \left ( \frac{2\pi k_B T(t,r)}{m_e}\right )^{\frac{3}{2}} \exp\left ( -\frac{m_ec^2}{k_B T(t,r)}\right) \text{,}
\label{Eq:pairsdensity}
\end{align}
where $h$ is the Planck constant.

\section{Computation of observed flux and spectrum}

The transparency of the shell occurs when its optical depth for Compton scattering reaches unity. For a shell, it writes \citep{RSV}:
\begin{align}
\tau (t,r,\phi_0)= \int_{r}^{R_{out}} \sigma_T n_c^{\pm} \Gamma (1-\beta \cos \phi) \frac{dR}{\cos(\phi)} \text{,}
\label{Eq:optic}
\end{align}
where $\phi$ is the laboratory angle between the radial direction and the four momentum of the photon, $\phi_0$ is that angle at the initial
radius from which the integration is performed, $\beta$ is the speed in units of the speed of light, $r$ is the radius of emission of the photon
and $R_{out}$ is the radius at which the photon leaves the shell. The integration has to be performed along the world line of a photon.

\cite{RSV} proposed two different approximations to compute the light-curves and the spectra: the fuzzy and the sharp photosphere ones.
In the sharp photosphere approximation, the energy contained in a small volume is assumed to be released instantly
at the time, radius and angle given by the condition $\tau(t,r,\phi)=1$. Then the laboratory energy $dE$ emitted in a laboratory solid
angle $d\Omega$ is equal to $dE = {3 \epsilon dV d\Omega} / ({8 \Lambda^4})$,
where $\Lambda=\Gamma (1-\beta \cos \phi)$ is the Doppler factor, 
$dV$ is the laboratory volume associated with the emission. 
In order to compute the light-curves, $dE$ is integrated over the photosphere for a given arrival time $t_a=t_e-\cos (\phi) (r_e/c)$, where $r_e$ and $t_e$ are the radial position and the laboratory
time of the emitting region, $t_a=0$ for a photon emitted at the origin.
The spectra are computed assuming that the energy is released with the Planck spectrum in the co-moving frame, with the co-moving temperature given by the co-moving energy density at the point
of emission. Then the intensity of radiation in the frequency range $d\nu_l$, solid angle $d\Omega_l$ is given by:
\begin{align}
I_{\nu} d\nu_l d\Omega_l = \int \frac{2 h}{ c^2} \frac{\nu_l^3 d\nu_l d\Omega_l dA}{\exp \left ( \frac{h\nu_l}{k T_l}\right )-1}\text{,}
\end{align}
where the quantities with index $l$ are measured in the laboratory frame, $d\Omega_l$ measures the
angular size of the detector. The integration over the surface $dA$ is performed to take into account the emitting area at the photosphere.
Due to the exponential dependence of the optical depth on the radial coordinate (see Eq.(\ref{Eq:pairsdensity}) and Eq.(\ref{Eq:optic})) the transition from the optically thick to the optically thin condition is indeed sharp. This fact justifies the sharp photosphere approximation in this problem.

The basis of the fuzzy photosphere approximation is the transfer equation for the specific intensity $I_\nu$ along the ray. Its formal solution is (see e.g Eq.(1.29) of \cite{rybicki}):
\begin{equation}
I_\nu = I_\nu(0)\exp(-\tau_\nu)+\int_0^{\tau_\nu} \exp(\tau_\nu-\tau_\nu^{'})S_\nu d\tau_\nu^{'}
\label{Eq:transferformal}
\end{equation}
where $S_\nu$ is the source function. At large optical depth, it is well known that the source function for scattering corresponds to a thermal
isotropic distribution of photon, with temperature $T(r,t)$. In addition, for energy dominated outflows, \cite{Beloborodov11} showed that coherent
scattering preserves the isotropy of the radiation field, together with the blackbody shape of the spectrum, like if radiation is propagating in vacuum. In fact, a freely propagating photon in an accelerating shell with $\Gamma\propto r$ does not change its angle in the co-moving frame. This later condition is actually used to obtain the approximate analytic
solution by \cite{bisno}. Since the radiation diffusion is negligible for accelerating shells \citep{RSV}, we can use the source function as a
thermal Planck function $S_\nu=B_\nu(T(t,r))$ and Eq.(\ref{Eq:transferformal}) can be integrated numerically. Then the flux at a given arrival time
is obtained by integration over the frequencies weighted by the surface of emission, while the total flux is obtained by additional integration
over the arrival time.

\section{Numerical results}

We now apply the approximate analytic solution of relativistic hydrodynamic equations obtained by \cite{bisno} to optically thick electron-positron plasma.
The parameters entering Eq.(\ref{Eq:epsilon}) and Eq.(\ref{Eq:gammaeq}) can be related to the total energy $E_0$ confined to a sphere of radius $R_0$ as follows:
\begin{align}
R_0 & =h_{shell}=3\sqrt{\frac{\delta_{fit}}{g_0}} \text{,}\\
E_0 & = 4 \pi g_0 \epsilon_{10} \times 1.3 h_{shell} \text{,}\\
\delta_{fit} & = 2.2\times 10^{-2} \text{.}
\end{align}

We perform the computation for $E_0=10^{54} \text{erg}$ and $R_0=10^8 \text{cm}$, corresponding to the initial temperature $k_B T_0=6.5 \text{MeV}$. Such parameters are typical within the Fireshell model of GRBs (see e.g. Ruffini et al. 2007)
We fixed the parameter $k$ to the value $1+\sqrt{3}/2$, as prescribed by \cite{bisno}.

The radial profiles of the laboratory energy density $\epsilon_l$, Lorentz factor and optical depth are displayed in Fig.\ref{Fig:profil} at
$ t = 250 R_0/c$. Since the solution of \cite{bisno} does not reproduce the separation between the front of the shell
and the light surface, the relative position of a photon inside the shell does not change substantially with time if it
propagates radially. The shell is photon thick \citep{RSV}: photons decouple because locally the pairs density decreases too fast
to sustain collisions.

The dependence of the optical depth on $\xi$ at a given laboratory time $t$ follows closely the variation of the co-moving energy density: from the
outer boundary toward
the centre it firstly increases and then decreases. This behaviour has to be contrasted with the one found by
\cite{RSV} for a simple shell profile of relativistic wind with finite duration:
the optical depth as a function of $\xi$ increases up to a saturation value from which it
stays constant up to the inner boundary of the outflow. This implies that the emission time $t_e$ on the line of sight is increasing
with the depth $\xi$ inside the shell.
In our more complex profile the outer and inner part of the shell become transparent before
the central part: there are two photospheres for a given laboratory time $t_e$. However, at a given arrival time $t_a$ there is only one photosphere:
the emission from the inner part of the shell arrives to the observer later.

\begin{figure}
\centering
\includegraphics[scale=0.7]{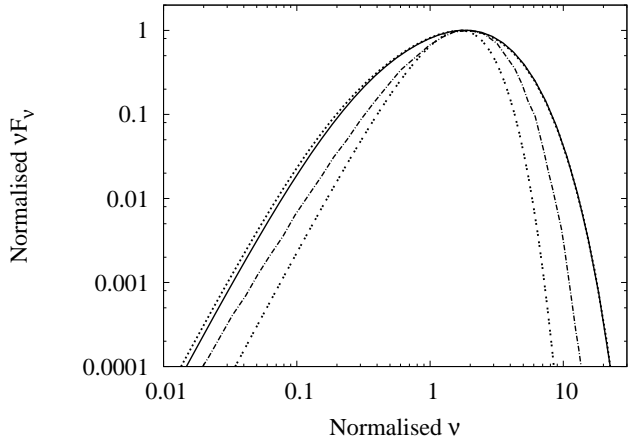}
\caption{Time integrated spectra: solid line-fuzzy approximation, dotted line-sharp approximation, dotted-dashed line-spectrum from Goodman (1986) and double-dashed line-Planck spectrum. For the sake of comparison these two last ones have been shifted to lower energy by a factor $2.3$.}
\label{Fig:spectra}
\end{figure}

\begin{figure}
\centering
\includegraphics[scale=0.35]{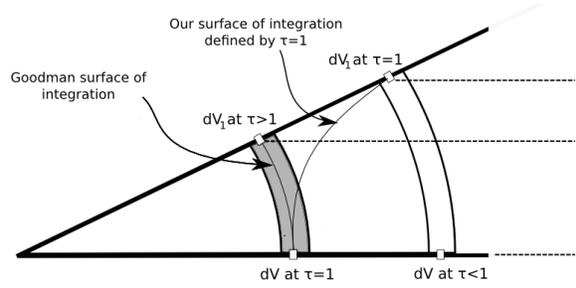}
\caption{Illustration of the difference in computation methods by Goodman and ours. The grey region corresponds to the shell at the laboratory time for which 
the optical depth of a photon propagating radially equals unity. It is the region in which the spectrum is computed by Goodman (1986). The shell is also represented at a larger
laboratory time with the curve linking $dV$ to $dV_1$ being a schematic representation of the $\tau=1$ surface at a given depth $\xi$.}
\label{Fig:scheme}
\end{figure}

Fig.\ref{Fig:spectra} displays the time-integrated spectra from sharp and fuzzy approximations, as well as the one obtained by
\cite{goodman} which has been shifted to lower energy by a factor $2.3$.
He obtained the observed spectrum by integration both over angles of emission and over the radial coordinate at a fixed laboratory time corresponding to
the moment of transparency on the line of sight, see the lower panel of Fig.\ref{Fig:profil}. This region of integration is shown in grey in
Fig.\ref{Fig:scheme}. In our computation the dynamics of the photosphere is taken into account explicitly: for each arrival time
the spectrum from the surface defined by $\tau(r,t,\phi_0)=1$ is computed. Eq.(\ref{Eq:optic}) shows that the optical depth increases with
$\phi_0$, so the points satisfying $\tau=1$ for a given $\xi$ are shifted to larger time and radius than the one on the line of sight. This surface
is displayed schematically by Fig.\ref{Fig:scheme}. Finally an integration is performed over arrival times.

This explains the two differences between our result and the one of Goodman. Firstly the time integrated spectrum is broader at low energy
because the observed temperature of a fluid element out of the line of sight is decreased in our computation. Secondly the peak energy is shifted
to lower energy because of the joint effect of increased volume of emission and decreased observed temperature out of the line of sight.

The low energy slopes are close in all cases
and are dominated by the high latitude emission. On the contrary the high energy part of the spectrum is dominated
by the photons emitted along the line of sight, for which the profile of temperatures plays an important role.

\begin{figure}
\centering
\includegraphics[scale=0.7]{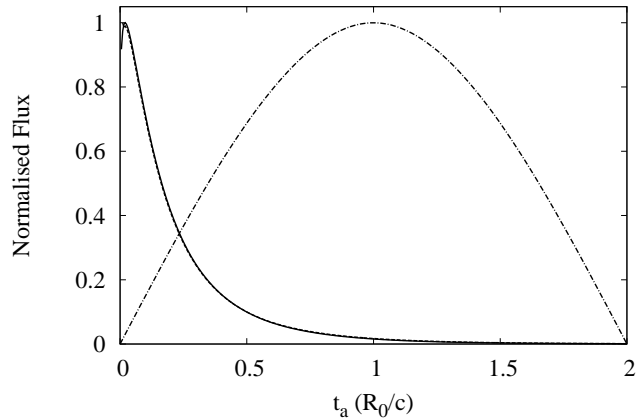}
\caption{Light-curves for fuzzy (continuous curve) and sharp (dashed curve) photosphere. They are nearly undistinguishable. For comparison the light-curve that would be obtained from an optically thin ball of radius $R_0$ uniformly filled by isotropic radiation is displayed by the dash-dotted line.}
\label{Fig:lc}
\end{figure}

The light-curves for sharp and fuzzy approximations are presented on Fig.\ref{Fig:lc}. Both approximations give close results, even if
the raising part is not resolved for the sharp photosphere approximation. Because of the narrow profile of the laboratory energy
density, the emission reaches its maximum and shortly later decreases: there is no plateau emission lasting the light-crossing time as reported
by \cite{RSV} for a finite wind. That is because they considered a different radial profile for the shell. Nevertheless the time needed to
emit $90\%$ of the energy is of the order of $R_0/c$.
 
\section{Discussion}

In our computation we used the simplifying assumption that the pairs recombine efficiently, so their number density is
given everywhere by Eq.(\ref{Eq:pairsdensity}). Nevertheless \cite{grimsrud} studied their recombination by considering the Boltzmann
equation in the case of a static and infinite wind. They showed that the pairs recombination process freezes out at the radius $R_{\pm}$, smaller
than $R_{ph}$. Above $R_\pm$, the co-moving pair density decreases proportionally to $r^{-3}$. The same effect is taken into account
in the Fireshell model, see \cite{ruffini99,ruffini00}. The ratio between the optical depth for Compton scattering and pairs recombination
process is:
\begin{align}
\frac{\tau_{\pm}}{\tau} = \frac{\sigma_\pm}{\sigma_T} \approx 0.8 \text{,}
\end{align}
where $\sigma_\pm$ is the cross section for the pairs recombination process. The co-moving temperature at $R_\pm$ can be found by solving
Eq.(54) of \cite{grimsrud}. We found $k_B T_\pm \sim 0.042 m_e c^2$, being close from the photosphere where the temperature is
$k_B T \sim 0.040 m_e c^2$ (see e.g. \cite{RSV}).

It implies that the optical depth increases when considering the freeze out of pairs recombination process, which leads to a small increase
in the value of $R_{ph}$. Secondly the optical depth for the pairs at $R_{ph}$ is given by:
\begin{align}
\tau_{\pm \gamma} (R_{ph}) = \tau (R_{ph}) \times \frac{n_\gamma}{n_{\pm}}  \gg 1 \text{,}
\end{align}
where the inequality holds because $\tau(R_{ph})=1$ by definition, and $n_\gamma \gg n_{\pm}$. It follows that even when the radiation streams
freely, the pairs are still strongly coupled to it and keep being accelerated by the radiative pressure: their Lorentz factor $\Gamma_\pm$
increases proportionally to $r$. For such outflow, \cite{beloborodov11a} showed that the isotropy of the radiation field is preserved in the
accelerating co-moving frame of pairs and that its temperature drops as $r^{-1}$. This means that the observed temperature is constant, and
no effect on the spectrum is expected.

The last point left to be discussed is the influence of the profile of temperatures within the shell. \cite{aksenov13} considered the decoupling
of photons from ultra-relativistic coasting winds with different profiles for the electron temperature. They showed that the spectral index
at low energy depends strongly on the chosen profile. Such kind of temperature dependence naturally arises when considering a realistic profile
for the expanding plasma. Nevertheless the position of a photon in the shell does not change substantially during the expansion below $R_{ph}$ for accelerated outflows, so no influence of the temperature profile on the photon Comptonization close to
the photosphere is expected in the spectrum. At larger radii when the radiation streams freely and crosses the shell,
the Compton parameter $y$ is much smaller than unity as the temperature is not relativistic and $\tau<1$, hence distortion of the spectrum
is small.

\section{Conclusion}

We have considered the analytical solution from \cite{bisno} for the spherical expansion
of a large amount of energy in a small volume and applied it to electron-positron plasma initially confined to a macroscopic volume.
Considering the dynamical evolution of the shell, we computed the flux and the energy distribution of the photospheric emission as seen
by a distant observer at rest in the laboratory frame. 

We found that the spectrum is broader than the Planck one and than the one of \cite{goodman} and shifted to lower energy, because
of the integration over impact parameters (or angles between the line of sight and the radial direction) in the dynamical photosphere.
The numerical results obtained by the sharp and fuzzy photosphere approximations coincide.

We additionally presented the light-curve from such event, showing that the maximum of the emission is reached in a short time scale compared to
the light-crossing time of the shell $R_0/c$. Then the flux decays sharply.

~

\section*{Acknowledgments}

We thank G.~S. Bisnovatyi-Kogan, as well as I.~A. Siutsou for useful discussions and comments. 

DB is supported by the Erasmus Mundus Joint Doctorate Program
by Grant Number 2011-1640 from the EACEA of the European Commission.

\begin{minipage}{150mm}

\appendix

\section{Approximate solution of the relativistic hydrodynamics equations}
\label{appendixA}

Starting from the laws of energy and momentum conservation, and by imposing the equation of state $\epsilon_c=3P$, one finds:
\begin{align}
\frac{\partial}{\partial t}\left ( 4 v \gamma^2 \epsilon \right ) + \frac{\partial}{\partial r}\left [ ( 4 v^2 \gamma^2 +1 ) \epsilon \right ]+\frac{8v^2\gamma^2\epsilon}{r} & = 0 \\
\frac{\partial}{\partial t}\left [( 4  \gamma^2 -1 )\epsilon \right ] + \frac{\partial}{\partial r}\left  ( 4 v \gamma^2 \epsilon \right )+\frac{8v\gamma^2\epsilon}{r} & = 0 \text{,}
\end{align}
where $v$ is such that $\gamma=(1-v^2)^{1/2}$.

\cite{bisno} makes the following change of variables $\xi=t-r$, $g=\gamma^2/r^2$, $\epsilon_1=\epsilon r^4$, $\epsilon_1=\epsilon_{10} \exp(-4\tau)$, $g=g_0 \exp(2\phi)$, $y_1=y/(2g)=r/(2\gamma^2)$, $x_1=\tau-(\phi/\sqrt{3})$ and $x_2=\tau+(\phi/\sqrt{3})$. Then the solution is extracted from the value of two functions $f$ and $\tilde{f}$ such that:
\begin{align}
f&=e^{-(1+\sqrt{3}/2)x_1}e^{-(1-\sqrt{3}/2)x_2} \xi \label{Eq:eq1}\\
 &= f_1(0) I_0(\sqrt{x_1 x_2}) + \int \limits_0^{x_2} \frac{df_1(x_2^{'})}{dx_2^{'}} I_0(\sqrt{x_1(x_2-x_2^{'})})dx_2^{'}+\int \limits_0^{x_1} \frac{df_2(x_1^{'})}{dx_1^{'}}I_0(\sqrt{(x_1-x_1^{'})x_2})dx_1^{'}\text{,} \\
\tilde{f}&=e^{-(1+\sqrt{3}/2)x_1}e^{-(1-\sqrt{3}/2)x_2} y_1 \\
& = \tilde{f}_1(0) I_0(\sqrt{x_1 x_2}) + \int \limits_0^{x_2} \frac{d\tilde{f}_1(x_2^{'})}{dx_2^{'}} I_0(\sqrt{x_1(x_2-x_2^{'})})dx_2^{'}+\int \limits_0^{x_1} \frac{d\tilde{f}_2(x_1^{'})}{dx_1^{'}}I_0(\sqrt{(x_1-x_1^{'})x_2})dx_1^{'}\label{Eq:eq4} \text{,}
\end{align}
where $I_0$s is the Bessel function and $f_1$, $f_2$, $\tilde{f}_1$ and $\tilde{f}_2$ are the following boundary functions
\begin{align}
f_1(x_2) & =(\xi_{a1}+\xi_a e^{-kx_2})e^{-(1-\sqrt{3}/2)x_2} \text{,} \\
\tilde{f}_1(x_2) & =(\xi_{a} y_a e^{-(1+k-\sqrt{3}/2 )x_2}\text{,} \\
f_2(x_1) & =\xi_{b} e^{-x_1}+\xi_{b1} e^{-(1+\sqrt{3}/2)x_1} \text{,} \\
\tilde{f}_2(x_1) & =(2-\sqrt{3})\xi_{b} e^{-x_1}\text{,}
\end{align}
where $\xi_a$, $\xi_{a1}$, $\xi_b$, and $\xi_{b1}$ are:
\begin{align}
\xi_a & = (2-\sqrt{3}) \frac{\xi_b}{\xi_a} \text{,} \\
\xi_{a1}-\xi_b & =\xi_b-\xi_a = \xi_0 \\
& = \sqrt{\frac{\delta_{fit}}{2g_0}}\sqrt{\frac{1}{2-\sqrt{3}}} \left (  1+\frac{2-\sqrt{3}}{2+\sqrt{3}}\frac{k-\sqrt{3}}{k} \right ) \text{,}
\end{align}
where finally we imposed $\xi_{a1}=0$ and $\xi_{0}= |\xi_{b1}|$.

The hydrodynamic profile is obtained in the following way: firstly, the equation $\xi-(1/y)-t=0$ links $x_1$ and $x_2$ at a given time by using Eq.(\ref{Eq:eq1}) to Eq.(\ref{Eq:eq1}), secondly the functions $f$ and $\tilde{f}$ are mapped for a given interval of $x_1$ (and correspondingly $x_2$), and finally the hydrodynamical profile is determined by using
\begin{align}
\epsilon & =\frac{2\epsilon_{10}g_0 f \tilde{f}}{\xi r^3}\\
\Gamma & = \sqrt{\frac{rf}{2 \xi \tilde{f}}} \text{,}
\end{align}
where $r$ and $\xi$ are computed using Eq.(\ref{Eq:eq1}).

\end{minipage}

\end{document}